\documentclass[a4paper,fleqn]{cas-dc}

\usepackage[numbers]{natbib}
\usepackage{bm}

\def\tsc#1{\csdef{#1}{\textsc{\lowercase{#1}}\xspace}}
\tsc{WGM}
\tsc{QE}

\begin{document}
\let\WriteBookmarks\relax
\def\floatpagepagefraction{1}
\def\textpagefraction{.001}

\shorttitle{Virtual Imaging Framework for 3D Quantitative Breast OAT}

\shortauthors{S. Park et~al.}

\title [mode = title]{A Virtual Imaging Framework for Three-Dimensional Quantitative Optoacoustic Tomography Using Stochastic Numerical Breast Phantoms}



\author[1]{Seonyeong Park}[type=editor, auid=000, bioid=1, orcid=0000-0002-4519-8693]
\fnmark[1] 
\ead{sp33@illinois.edu} 
\credit{Conceptualization of this study, Meth-odology, Software, Data curation, Writing - original draft} 
\affiliation[1]{organization={Department of Bioengineering, University of Illinois Urbana-Champaign},
            addressline={1406 West Green Street}, 
            city={Urbana},
            postcode={61801}, 
            state={IL},
            country={United States}}

\author[1]{Gangwon Jeong}[type=editor, auid=000,bioid=1, orcid=0000-0002-5142-2559]
\ead{gangwon2@illinois.edu} 
\credit{Software} 

\author[2,3]{Umberto Villa}[type=editor, auid=000, bioid=1, orcid=0000-0002-5142-2559]
\ead{uvilla@austin.utexas.edu} 
\credit{Software, Writing - review \& editing}
\affiliation[2]{organization={Department of Biomedical Engineering, The University of Texas at Austin},
            addressline={107 W Dean Keeton St}, 
            city={Austin},
            postcode={78712}, 
            state={TX},
            country={United States}}
\affiliation[3]{organization={Oden Institute for Computational Engineering and Sciences, The University of Texas at Austin},
            addressline={201 E. 24th Street}, 
            city={Austin},
            postcode={78712}, 
            state={TX},
            country={United States}}

\author[1]{Mark A. Anastasio}[type=editor, auid=000, bioid=1, orcid=0000-0002-3192-4172]
\cormark[1] 
\ead{maa@illinois.edu} 
\ead[url]{https://anastasio.bioengineering.illinois.edu/} 
\credit{Conceptualization of this study, Supervision, Writing - review \& editing}
\cortext[1]{Corresponding author} 



\begin{abstract}
Optoacoustic tomography (OAT) is a promising modality for breast cancer diagnosis because tumor angiogenesis and, potentially, hypoxia can be visualized using quantitative OAT (qOAT) techniques. Clinically meaningful inference generally requires accurate image reconstruction, which depends on measurement quality and the imager design. Virtual imaging offers a cost-effective alternative to experimental prototyping for system design evaluation and supports computational method development. This work presents a comprehensive virtual imaging framework for breast qOAT, extending a stochastic numerical breast phantom (NBP) generator by incorporating skin tone variation and both benign and malignant lesions. It enables end-to-end simulation, modeling transducer spatial and acousto-electric impulse responses. Its utility is demonstrated through a case study comparing two system designs. A total of 1,020 NBPs and associated measurement data have been made publicly available to accelerate research in optoacoustic and optical imaging. The framework provides a versatile platform for advancing computational methods and guiding system design optimization. 
\end{abstract}





\maketitle

\section{Introduction}
\label{SEC1:introduction}
Three-dimensional (3D) optoacoustic tomography (OA-T), also known as photoacoustic computed tomography (PACT), enables volumetric imaging of the absorbed optical energy density in biological tissues using short, non-ionizing laser pulses, typically at near-infrared (NIR) wavelengths~\cite{Oraevsky2001, Oraevsky2018, louisa, Lin2018, Park2023, Park2022}. Optical illumination induces transient thermoelastic expansion, which generates initial pressure distributions through the photoacoustic effect~\cite{Park2023, Lin2018}. Ultrasonic transducers surrounding the tissue detect the resulting acoustic waves, which propagates through the tissue~\cite{Oraevsky2001, Oraevsky2018, Lin2018, Lou2017}. These measurements are subsequently used for image reconstruction~\cite{Poudel2019, Chen2024, Chen2025, Jeong2025}. On a macroscopic scale, OAT achieves higher spatial resolution at greater tissue depths than purely optical imaging techniques because acoustic waves scatter substantially less than light in biological tissues~\cite{Oraevsky2018, Lin2018, Yao2016}. In addition, OAT exploits strong optical contrast from the spectral absorption of endogenous chromophores, primarily oxy- and deoxy-hemoglobin, yielding greater contrast than conventional ultrasound imaging~\cite{Oraevsky2018, Park2023, Park2022}.

In OAT, the induced initial pressure distribution is reconstructed by solving an acoustic source inverse problem using measured data~\cite{Poudel2019, Chen2024, Chen2025, Jeong2025, qpact}. This enables visualization of blood vasculature, which is highly absorptive due to its rich hemoglobin content. In principle, functional parameters, such as blood oxygen saturation, and molecular composition can be estimated from acoustic measurements at multiple light excitation wavelengths using techniques known as quantitative OAT (qOAT), also referred to as quantitative PACT~\cite{Park2022, Park2023, qpact, Cam2024}. These estimates are obtained by solving a multiphysics inverse problem that involves both photon transport and acoustic wave propagation~\cite{qpact, Cam2024}. By delineating vascular anatomy and quantifying functional parameters, optoacoustic imaging provides both structural and functional insights. These capabilities make OAT particularly well suited for diagnosis of breast cancer characterized by features such as tumor angiogenesis and hypoxia~\cite{Toi2017, Oraevsky2018, Neuschler2017, Dogan2019, Oraevsky2001, Lin2018, Park2023}.

Accurate and robust image reconstruction is generally important for extracting clinically meaningful information from OAT images, particularly under diverse clinical conditions. In general, the quality of reconstructed images is fundamentally limited by the quality of measured data, which depends on the design of the OAT imaging system. Unlike many other imaging modalities, 3D OAT imagers offer considerable design flexibility, as each consists of a light delivery subsystem and an acoustic data acquisition subsystem~\cite{Oraevsky2018, Lin2018, Toi2017}. In the acoustic data acquisition subsystem, detector characteristics such as sensitivity and bandwidth determine the achievable signal-to-noise ratio (SNR) and spatial resolution~\cite{Yao2016, Oraevsky2018}. Measurement geometry, defined by the spatial arrangement of detectors, and temporal sampling, determined by the acquisition rate, together establish the data completeness, a critical factor for image reconstruction~\cite{Chen2024, Poudel2019}. The configuration of the light delivery subsystem directly determines the initial pressure distribution generated within the tissue, and thus, data quality~\cite{Park2022}. The broad range of possible design choices, each of which may significantly affect image quality, necessitates careful evaluation and optimization. This process has been challenging due to the need to construct physical prototypes and well-characterized phantoms, acquire experimental data, and perform comparative validations~\cite{Park2023, Park2023_skin, Chen2025}.

To address these challenges, virtual imaging provides a cost-effective and ethical alternative for systematically exploring the extensive parameter space inherent in OAT system design, facilitating efficient optimization at a fraction of the cost and effort required for experimental imaging~\cite{Park2023, Park2023_skin, Chen2025, Li2022}. Virtual imaging enables the generation of simulated measurement data with complete control over physical factors that affect data quality, which enables the analysis of competing imaging system designs. Through the specification of the to-be-imaged object, virtual imaging also allows for systematic simulation of a wide range of physiological and pathological conditions, supporting diagnostic task-based assessment of image quality~\cite{Abadi2020, victre, Myers2023, Abadi2024}. However, the practical value of these advantages depends critically on how realistically the virtual objects can be modeled in physiological, optical, and acoustic terms, as well as on the fidelity with which the relevant imaging physics are represented.

Previous virtual imaging studies of OAT have often relied on oversimplified numerical phantoms~\cite{fdasimple, Fadden2018, Sowmiya2017}. Validation of computational methods or imaging system designs employing such phantoms may result in misleading conclusions. In several studies~\cite{Lou2017, Ma2020}, anatomically realistic datasets of objects were generated by segmenting clinical images acquired from other modalities. However, the number of images in the reported datasets is generally limited. Additionally, in most previous studies, optical properties were directly assigned at a few preselected wavelengths without modeling their dependence on chromophore concentrations, limiting their utility for functional imaging~\cite{fdasimple, Fadden2018, Sowmiya2017, Lou2017, Ma2020, Bao2021}. To overcome these limitations and ensure that virtual imaging studies are applicable in practice, virtual objects must reflect anatomical and physiological variability and model tissue optics by linking chromophore concentrations to optical absorption coefficients~\cite{Park2023, Cam2024}. Stochastic object models~\cite{zhou2022learning} represent a means to capture the diversity needed for robust method development and system optimization~\cite{Park2023, Cam2024, Chen2024, Yang2025}. Flexible wavelength selection is also desirable for greater versatility.

To fully benefit from virtual imaging, the computational model must closely approximate the physics of real-world imaging. Without high model fidelity, insights gained from simulation may not translate to real-world performance on experimental data. Light sources are frequently approximated with alternative setups because matching experimental illumination conditions is not always feasible~\cite{Park2023, Lou2017}, suggesting opportunities to extend current simulation capabilities. Additionally, incorporating detailed transducer response modeling and accurate measurement geometry into large-scale 3D simulations is impractical due to computational constraints~\cite{k-wave}. Implementation-level improvements to current simulation tools may enable scalable simulations and enhance acoustic modeling fidelity.

A software tool for generating 3D numerical breast phantoms (NBPs) for use in OAT imaging, referred to as the stochastic optoacoustic NBP (SOA-NBP)~\cite{Park2023}, has been developed. Phantoms generated using SOA-NBP exhibit stochastic, anatomically and optoacoustically realistic spatial distributions of breast tissue and malignant lesions, enabling virtual imaging studies that reflect clinically relevant variability. By incorporating a wavelength-dependent tissue optics model parameterized by underlying physiological quantities, the framework enables the generation of phantoms at arbitrary wavelengths within the NIR range, which is required for their use in qOAT virtual imaging studies. Eighty-four NBP datasets have been released, each providing an anatomical map as well as functional, optical, and acoustic property maps that collectively define the phantom. Each dataset also includes the corresponding optical fluence maps, initial pressure maps, and pressure measurements, all simulated at three wavelengths using a virtual imaging system configuration that emulates an existing experimental setup~\cite{louisa}. These datasets have already been employed in studies within the optoascoustic (photoacoustic) and optical imaging communities~\cite{Chen2024, Chen2025, Yang2025, Cam2024, Park2023_skin, Hossain_2025}.

In this work, a comprehensive virtual imaging framework for qOAT of the breast is proposed. This framework builds on the previously developed SOA-NBP software~\cite{soa-nbp}, which stochastically generates 3D NBPs. The proposed framework extends the SOA-NBP by incorporating a detailed skin model that accounts for skin-tone-dependent optical absorption and by introducing models of benign lesions in addition to malignant tumors. These enhancements enable the framework to represent the anatomical and physiological diversity of the target population and to guide system designs toward robust performance across such variability. The simulation components have also been advanced to more faithfully simulate the entire imaging process. Specifically, the framework incorporates the transducers' spatial impulse response (SIR) in large-scale 3D simulations and their acousto-electric impulse response (EIR) modeled from experimental data. These extensions establish a versatile virtual imaging environment that can facilitate a variety of research applications. The datasets generated in this study using the proposed framework have been made publicly available (see Section~\ref{data} for details). 

The contributions of this work are fourfold. First, an extensive review of virtual imaging in 3D OAT, including commonly employed tools, is presented. Second, building upon previous work, a publicly available collection of 1,020 anatomically and optoacoustically realistic NBPs is provided, together with the corresponding optical fluence maps, initial pressure maps, and pressure measurements to support and accelerate research within the broader optical and optoacoustic imaging communities. The diversity spans breast density types, multiple skin tones, and lesion phenotypes. This dataset is particularly well suited for the development and evaluation of data-driven techniques beyond image reconstruction, including a wide range of image processing and image analysis tasks. Third, a reproducible and modular simulation framework, incorporating extended functionalities of existing tools, provides a practical foundation for modeling realistic OAT imaging scenarios. Fourth, the utility of the proposed virtual imaging framework is demonstrated through a system optimization case study.

The remainder of the paper is organized as follows. Section~\ref{SEC2:background} provides reviews on SOA-NBP and virtual imaging in OAT. The proposed virtual imaging framework is detailed in Section~\ref{SEC3:vi_framework}. Section~\ref{SEC4:examples} presents examples NBPs and the corresponding OAT images. An application of the framework for evaluating example OAT system designs is demonstrated in Section~\ref{SEC5:case_study}. The paper concludes in Section~\ref{SEC6:conclusion}.

\section{Background}
\label{SEC2:background}
SOA-NBP, a key component in the considered 3D virtual imaging framework, is reviewed in Section~\ref{SEC2.1:soa-nbp}. Section~\ref{SEC2.2:physics_fowd_model_of_oat} describes the governing physics and forward model of OAT. A brief review of an existing virtual imaging framework for optoacoustic imaging is provided in Section~\ref{SEC2.3:vif_oa_imaging}.

\subsection{SOA-NBP: Stochastic Optoacoustic Numerical Breast Phantom}
\label{SEC2.1:soa-nbp}
SOA-NBP~\cite{Park2023} is a software tool for generating ensembles of anatomically and physiologically realistic 3D NBPs for use in OAT imaging. This tool was developed by substantially modifying and extending the phantom generation software originally created for digital mammography and digital breast tomosynthesis as part of the U.S. FDA's virtual imaging clinical trials for regulatory evaluation (VICTRE) project~\cite{victre}. As in the original FDA phantom, breast shape, size, and tissue composition are user-controllable. Four breast types, as defined in the breast imaging reporting and data system (BI-RADS)~\cite{bi-rads}, are supported: (A) almost entirely fatty breasts, (B) breasts with scattered areas of fibroglandular density, (C) heterogeneously dense breasts, and (D) extremely dense breasts. The NBP consists of various tissue types: fat, skin, glandular tissue, nipple, muscle, ligament, terminal duct lobular unit, duct, artery, and vein. 

SOA-NBP addresses the specific requirements of OAT virtual imaging by incorporating an improved blood vasculature model and an advanced malignant lesion model. The lesion model explicitly represents internal tumor heterogeneity, including a viable tumor cell (VTC) region, a necrotic core, and a peritumoral angiogenesis region. To ensure physiological consistency, the optical absorption coefficient is computed based on the local concentrations of major chromophores: oxy- and deoxy-hemoglobin, water, fat, and melanosomes. Specifically, the optical absorption coefficient in each tissue is calculated as a weighted sum of the absorption coefficients of these pure chromophores, with the weights corresponding to their respective volume fractions~\cite{Jacques2013, Prahl}. This approach supports flexible generation of tissue optical property maps across the NIR range of 700~nm to 1100~nm, which is commonly used in OAT~\cite{louisa, Oraevsky2001, Lin2018, Toi2017, Oraevsky2018}. Furthermore, the blood oxygen saturation distribution is modeled to represent smooth, biologically realistic transitions between tissue regions. 

Each NBP produced using SOA-NBP consists of anatomical, functional, optical, and acoustic property distributions. The functional properties correspond to total hemoglobin concentration $c_{\text{tHb, blood}}$ ($\mu$M), blood oxygen saturation $s$ (\%), and the volume fractions of blood $f_b$, water $f_w$, fat $f_f$, and melanosome $f_m$ (\%). The optical properties include optical absorption coefficient $\mu_a$ (mm$^{-1}$), scattering coefficient $\mu_s$ (mm$^{-1}$), scattering anisotropy $g$, and refractive index $n$. The acoustic properties are sound speed $c$ (mm/$\mu$s), density $\rho$ (g/mm$^3$), and attenuation coefficient $\alpha_0$ (dB/MHz$^y$) with power law exponent $y$. All properties are assigned to each tissue and are stochastically prescribed within physiologically realistic ranges determined through a comprehensive literature survey~\cite{Park2023}.

NBP data sets produced using SOA-NBP have been deployed in studies that involve the estimation of the initial pressure distribution by solving the acoustic inverse problem~\cite{Chen2024, Chen2025, Yang2025}, joint estimation of the initial pressure and sound speed distributions~\cite{Jeong2025, Suhonen2025}, estimation of functional quantities from multi-wavelength acoustic measurements through multiphysics inversion~\cite{Cam2024}, and the comparison of light delivery subsystem designs in a 3D OAT breast system~\cite{Park2023_skin}. Specifically, the datasets have supported the development of learning-based methods~\cite{Cam2024, Chen2024} as well as the validation of existing and newly proposed methods~\cite{Jeong2025, Yang2025, Chen2025, Cam2024}. Notably, a method developed using this dataset demonstrated robust generalizability to experimental data~\cite{Chen2024}. These applications demonstrate the practical utility of the datasets for the optoacoustic (photoacoustic) imaging community. Furthermore, the optical property distributions included in these datasets have also been utilized in research on diffuse optical tomography beyond OAT~\cite{Hossain_2025}, highlighting their versatility and value for the broader optical imaging community.

\subsection{Physics and Forward Model of OAT}
\label{SEC2.2:physics_fowd_model_of_oat}
OAT virtual imaging refers to the computational simulation of the OAT forward process using numerical object models. This process consists of two stages: (1) simulation of the optoacoustically induced initial pressure field, and (2) simulation of its propagation through tissue and detection by ultrasonic transducers surrounding the tissue~\cite{Park2023, qpact}. 

In the first stage, the initial pressure distribution at position $\bm{r}\in\mathbb{R}^3$ and time $t=0$, induced by optical illumination at wavelength $\lambda$, is given by: 
\begin{equation}
\label{eq:1}
\begin{aligned}
p(\bm{r}, t, \lambda)|_{t=0} &= p_0(\bm{r}, \lambda) = \Gamma(\bm{r})\,A(\bm{r}, \lambda) \\
&= \Gamma(\bm{r})\,\mu_a(\bm{r}, \lambda)\,\phi(\bm{r}, \lambda),
\end{aligned}
\end{equation}
where $A(\bm{r}, \lambda)$ is the absorbed optical energy density, $\phi(\bm{r}, \lambda)$ is the optical fluence, and $\Gamma(\bm{r})$ is the dimensionless Gr\"{u}neisen parameter quantifying the conversion efficiency from absorbed optical energy to initial pressure via thermoelastic expansion. For soft tissue, $\Gamma(\bm{r})$ is typically assumed to be constant and is often set to 1 for simplicity~\cite{fdasimple,Park2023}. In this case, $p_0(\bm{r}, \lambda)$ is equivalent to $A(\bm{r}, \lambda)$. 

In OAT, light transport is described by the radiative transfer equation (RTE), assuming conservation of photon energy during collisions and a constant refractive index within the medium~\cite{qpact}. Because direct solutions of the RTE are computationally demanding, two approaches are commonly used for practical computation of $\phi(\bm{r}, \lambda)$: the diffusion approximation (DA)~\cite{qpact} and the Monte Carlo (MC) method~\cite{Wang1995}. In the DA, the RTE is approximated by the diffusion equation under the assumption that the medium's reduced scattering coefficient, $\mu_s' = \mu_s(1 - g)$, greatly exceeds its absorption coefficient, $\mu_a$~\cite{qpact}. This assumption does not hold in media containing non-scattering regions, such as water used for acoustic coupling in OAT imaging, resulting in inaccurate estimates. In contrast, the MC method, widely regarded as the gold standard, accurately models photon propagation in scattering and absorbing media without relying on the simplifying assumptions required by the DA~\cite{qpact}. MC simulations use random sampling to stochastically model the trajectories of numerous photons and estimate $\phi(\bm{r}, \lambda)$ from the resulting photon distribution. 

In the second stage, acoustic wave propagation is governed by the acoustic wave equation, and the pressure at detection locations is obtained by solving this equation. Acoustic attenuation in biological soft tissues is typically described by a frequency power law~\cite{Szabo2013}:
\begin{equation}
\label{eq:2}
\alpha(\bm{r}, \omega) = \alpha_0(\bm{r})\omega^y,
\end{equation}
where $\alpha$ is the acoustic attenuation coefficient, $\alpha_0$ is the frequency-independent attenuation coefficient, $\omega$ is the angular temporal frequency, and $y$ is the power law exponent. For practical numerical implementations, $y$ is typically assumed to be constant~\cite{Treeby2010}. Modeling soft tissue as an acoustically heterogeneous, lossy medium, the optoacoustically induced pressure wavefield $p(\bm{r}, t, \lambda)$, at position $\bm{r} \in \mathbb{R}^3$, time $t \geq 0$, and optical excitation wavelength $\lambda$, can be described by a coupled system of first-order partial differential equations~\cite{Tabei2002, Treeby2010}:
\begin{subequations}
\label{eq:3}
\begin{align}
\frac{\partial}{\partial t} \bm{u}(\bm{r}, t) &= -\frac{1}{\rho_0(\bm{r})} \nabla p(\bm{r}, t, \lambda), \\
\frac{\partial}{\partial t} \rho(\bm{r}, t) &= -\rho_0(\bm{r}) \nabla \cdot \bm{u}(\bm{r}, t), \\
p(\bm{r}, t, \lambda) &= c_0(\bm{r})^2 \bigg\{1 - \mu(\bm{r})\frac{\partial}{\partial t}\big(-\!\nabla^2\big)^{y/2-1} \notag \\ 
&\qquad\quad
- \eta(\bm{r})\big(-\!\nabla^2\big)^{(y-1)/2}\bigg\}\rho(\bm{r}, t),
\end{align}
\end{subequations}
subject to the initial conditions in Eq. \eqref{eq:1} and $\bm{u}(\bm{r}, t)|_{t=0} = 0$. Here, $\bm{u}(\bm{r}, t) \equiv \big(u^1(\bm{r}, t), u^2(\bm{r}, t), u^3(\bm{r}, t)\big)$ is a vector-valued function that represents the acoustic particle velocity field, $\mu(\bm{r}) = -2\alpha_0(\bm{r})c_0(\bm{r})^{y-1}$ denotes the acoustic absorption coefficient, and $\eta(\bm{r}) = 2\alpha_0(\bm{r}) c_0(\bm{r})^{y}\tan(\frac{\pi y}{2})$ is the dispersion proportionality coefficient. The $k$-space pseudo-spectral time-domain method is commonly employed to discretize and solve these coupled equations, as it offers high accuracy, low numerical dispersion, stability, and computational efficiency~\cite{Tabei2002}.

\subsection{Existing Virtual Imaging Frameworks for Optoacoustic Imaging}
\label{SEC2.3:vif_oa_imaging}
For virtual imaging studies in optoacoustic imaging, several tools are available for simulating photon transport and acoustic wave propagation (see Appendix~\ref{APPENDIX:A}). The most widely used are MCX~\cite{mcx1, mcx2}, a GPU-accelerated photon transport simulator based on the MC method, and k-Wave~\cite{k-wave}, an acoustic wave simulator that solves the acoustic wave equation using the above mentioned $k$-space pseudo-spectral time-domain method. To streamline simulation workflows, a toolkit called SIMPA~\cite{simpa} has been developed. SIMPA is a Python-based adapter that operates at the configuration layer, providing a user-friendly infrastructure to connect simulation modules and manage data flow between them. It also enables the creation of simple tissue phantoms with shapes such as layers, spheres, elliptical tubes, cuboids, parallelepipeds, and vessel trees, and supports system configuration for commercial imagers, specifically the MSOT Acuity Echo, InVision 256-TF~\cite{Janek2025}, and RSOM Xplorer P50 from iThera Medical (Munich, Germany). The authors who developed SIMPA have also applied it in comparative studies of two-dimensional (2D) OAT image reconstruction methods~\cite{Janek2025}, demonstrating its utility for image quality assessment studies.

\section{Virtual Imaging Framework for 3D qOAT of the Breast}
\label{SEC3:vi_framework}
\begin{figure}
\centering
\includegraphics[width=1\columnwidth]{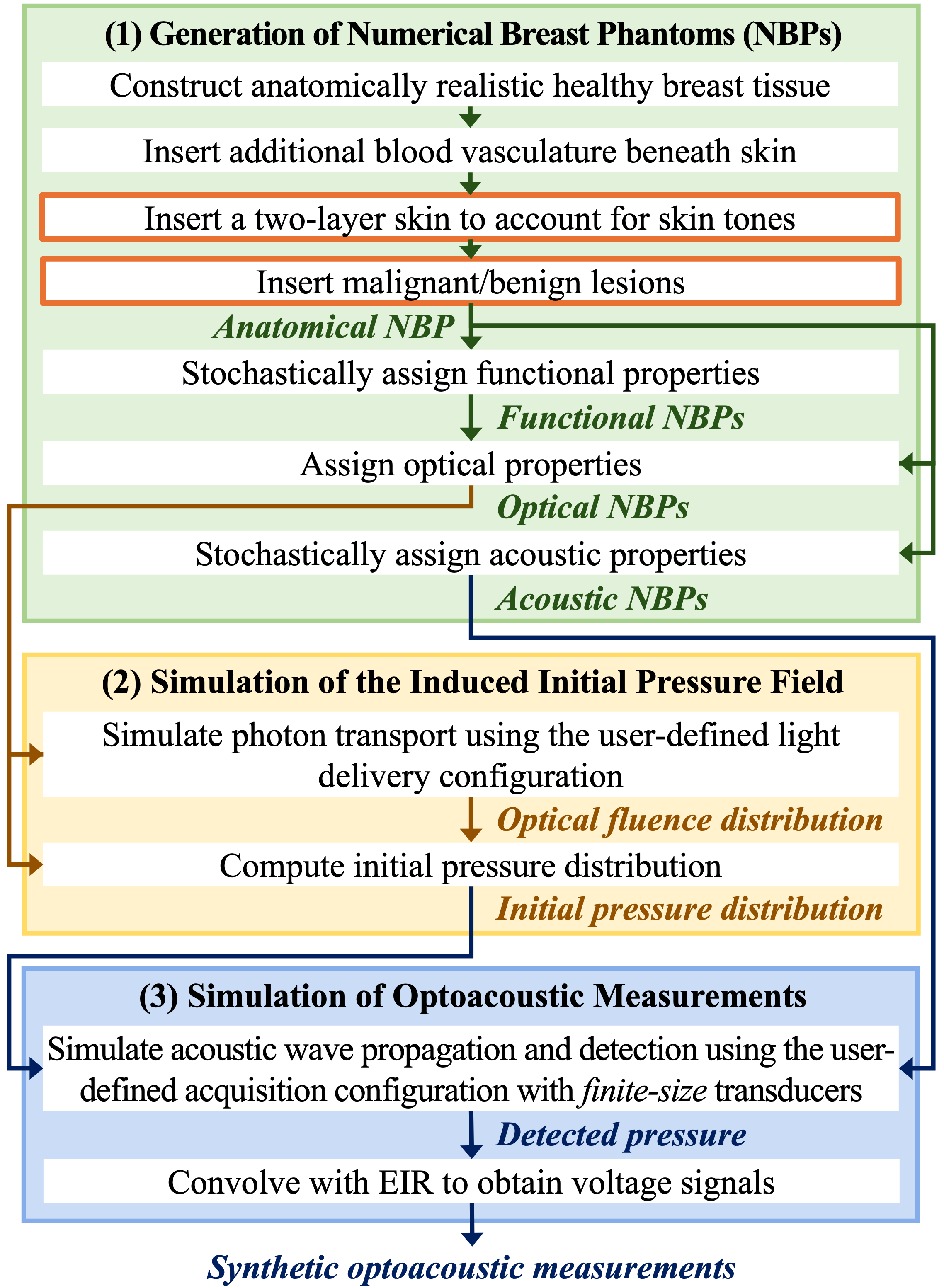}
\caption{Virtual imaging framework for 3D qOAT of the breast. Anatomical, functional, optical, and acoustic NBPs are generated using the SOA-NBP tool~\cite{Park2023}, with extended features indicated by orange-outlined boxes. The generated optical NBPs are used to compute the optoacoustically induced initial pressure field. The resulting distribution and the acoustic NBPs are then used as inputs to simulate acoustic wave propagation and optoacoustic measurements, accounting for the impulse responses of the transducers.}
\label{FIG:1}
\end{figure}

This section presents a comprehensive virtual imaging framework for 3D qOAT of the breast. As illustrated in Fig.~\ref{FIG:1}, the framework consists of three primary steps: (1) generation of optoacoustic NBPs, (2) simulation of the optoacoustically induced initial pressure field via photon transport, and (3) simulation of optoacoustic measurements through acoustic wave propagation. 

Optoacoustic NBPs were generated using the previously proposed SOA-NBP tool, which produces anatomically realistic healthy breast tissue maps with embedded subcutaneous blood vasculature. In this study, the tool was extended to include a two-layer skin model that accounts for skin tone variations, as well as an expanded set of lesion types that can optionally be inserted. Functional, optical, and acoustic properties were then stochastically assigned to each tissue type yielding the complete NBP set. The simulation of the optoacoustically induced initial pressure field involves computing the optical fluence distribution and then deriving the initial pressure distribution based on the optical NBPs. The fluence computation was implemented with additional light source customization. Optoacoustic measurements were obtained by simulating acoustic wave propagation and detection, followed by conversion of the detected pressure signals into voltage signals. To faithfully replicate imaging physics under real-world system configurations, both the SIR and EIR of transducers were incorporated.

Details of the extended SOA-NBP are provided in Section~\ref{SEC3.1:extended_soa-nbp}. The two simulation steps are described in Sections~\ref{SEC3.2:photoacoustic_effect} and~\ref{SEC3.3:optoacoustic_measurements}, respectively, with emphasis on the extensions introduced in this study.

\subsection{Extended SOA-NBP}
\label{SEC3.1:extended_soa-nbp}
The previous SOA-NBP implementation~\cite{Park2023} focused on a single skin tone and malignant tumors, without accounting for variability across other skin tones or lesion types. To broaden its applicability, the extended version models various skin tones and additionally incorporated two benign lesion models. These features are described in Sections~\ref{SEC3.1.1:skin_tones} and~\ref{SEC3.1.2:lesion_types}, respectively.

\subsubsection{Modeling Skin Tones}
\label{SEC3.1.1:skin_tones}
Melanin is a chromophore that exhibits a higher optical absorption coefficient than oxy- and deoxy-hemoglobin in the NIR range of 700 to 1100~nm~\cite{Jacques2013}. Melanin is distributed exclusively in the epidermis, the most superficial layer of the skin, and its concentration determines skin tone~\cite{Jacques2013}. Variations in skin tone, therefore, directly affect the amount of light absorbed in the superficial skin layer during an OAT scan~\cite{Jacques1998, Park2023}. Darker skin tones have higher melanin concentrations, resulting in a reduction of the amount of light that penetrates to deeper tissues~\cite{Park2023_skin, Else2023, Rasquinha2024}. This reduction in optical fluence compounds the depth-dependent optical attenuation, further decreasing the optoacoustic contrast in deeper regions and thereby lowering the SNR. Despite this important effect, the extent to which skin tone affects OAT imaging remains relatively unexplored. Only recently have a few studies~\cite{Park2023_skin, Else2023, Rasquinha2024} begun to investigate this effect. Incorporating physiologically realistic skin tone variability into virtual imaging frameworks enables the generation of synthetic optoacoustic data that capture skin pigmentation differences, thereby supporting the systematic investigation of skin tone's impact on signal detectability and diagnostic performance. 

To represent skin tone variability in SOA-NBP, a two-layer skin model was introduced, consisting of a dermis-only layer and an epidermis-included layer. Total skin thickness, as measured in mammographic studies, ranges from 0.5 to 3.1~mm, including both the epidermis and dermis~\cite{Ulger2008, Huang2008}, with the epidermis accounting for approximately 3.7--16.8\% of the total thickness~\cite{Huang2008}. Because the spatial resolution currently achievable with OAT (a few hundred microns)~\cite{louisa, Lin2018, Toi2017} is insufficient to resolve the epidermis, the epidermis-included layer in the anatomical phantom was defined as a one-voxel-thick region along the outer skin surface encompassing the entire epidermis and part of the dermis. The thicknesses of the total skin and epidermis are user-configurable. In this study, they were set to 1.5~mm and 0.02~mm, respectively, as an example. Voxels labeled as skin in the previous SOA-NBP were assigned to the dermis-only layer. The epidermis-included layer was defined as the set of dermis voxels directly adjacent to the background in any of the $x$, $y$, or $z$ directions. The definitions of newly introduced or reassigned tissue type labels are provided in Appendix~\ref{APPENDIX:B}.

Skin tone is commonly classified using the Fitzpatrick phototype scale~\cite{Fitzpatrick1988}, which ranges from type I (always burns, never tans) to type VI (never burns), based on the skin's response to ultraviolet exposure. In phototype I, melanin contributes minimally to optical absorption, resulting in limited contrast in OAT images, whereas phototype VI contains four to eight times more melanin, leading to substantially higher optical absorption within the epidermis~\cite{Tseng2009, Jacques1998}. Melanin concentration is represented by $f_{m}$, the volume fraction of melanosomes, which are the organelles that synthesize melanin. In SOA-NBP, $f_{m}$ is defined as one of the functional properties, together with the volume fractions of blood $f_b$, water $f_w$, and fat $f_f$, total hemoglobin concentration $c_{\text{tHb, blood}}$, and blood oxygen saturation $s$. For phototypes I to VI, $f_{m}$ was sampled from uniform distributions within the ranges specified in Table~\ref{tbl1}~\cite{Jacques1998}. Because melanosomes are absent in the dermis, the dermis-only layer was assigned $f_{m}=0$. For the epidermis-included layer, the effective $f_{m}$ was calculated as the arithmetic average based on the configured epidermal thickness and voxel size. The remaining functional, optical, and acoustic properties were assigned to each tissue type as described in~\cite{Park2023}. 

\begin{table}[width=.8\linewidth,cols=4,pos=h]
\caption{Melanosome volume fraction $f_m$ in epidermis~\cite{Jacques1998}}\label{tbl1}
\begin{tabular*}{\tblwidth}{@{}LL@{}}
\toprule
Fitzpatrick skin phototype & $f_{m}$ range (\%) \\ 
\midrule
I       & [1.3, 3.8) \\
II      & [3.8, 6.3) \\
III     & [6.3, 9.3) \\
IV      & [9.3, 14.3) \\
V       & [14.3, 21.3) \\
VI      & [21.3, 30.3] \\
\bottomrule
\end{tabular*}
\end{table}

\subsubsection{Modeling Lesion Types}
\label{SEC3.1.2:lesion_types}
\begin{figure}
\centering
\includegraphics[width=0.95\columnwidth]{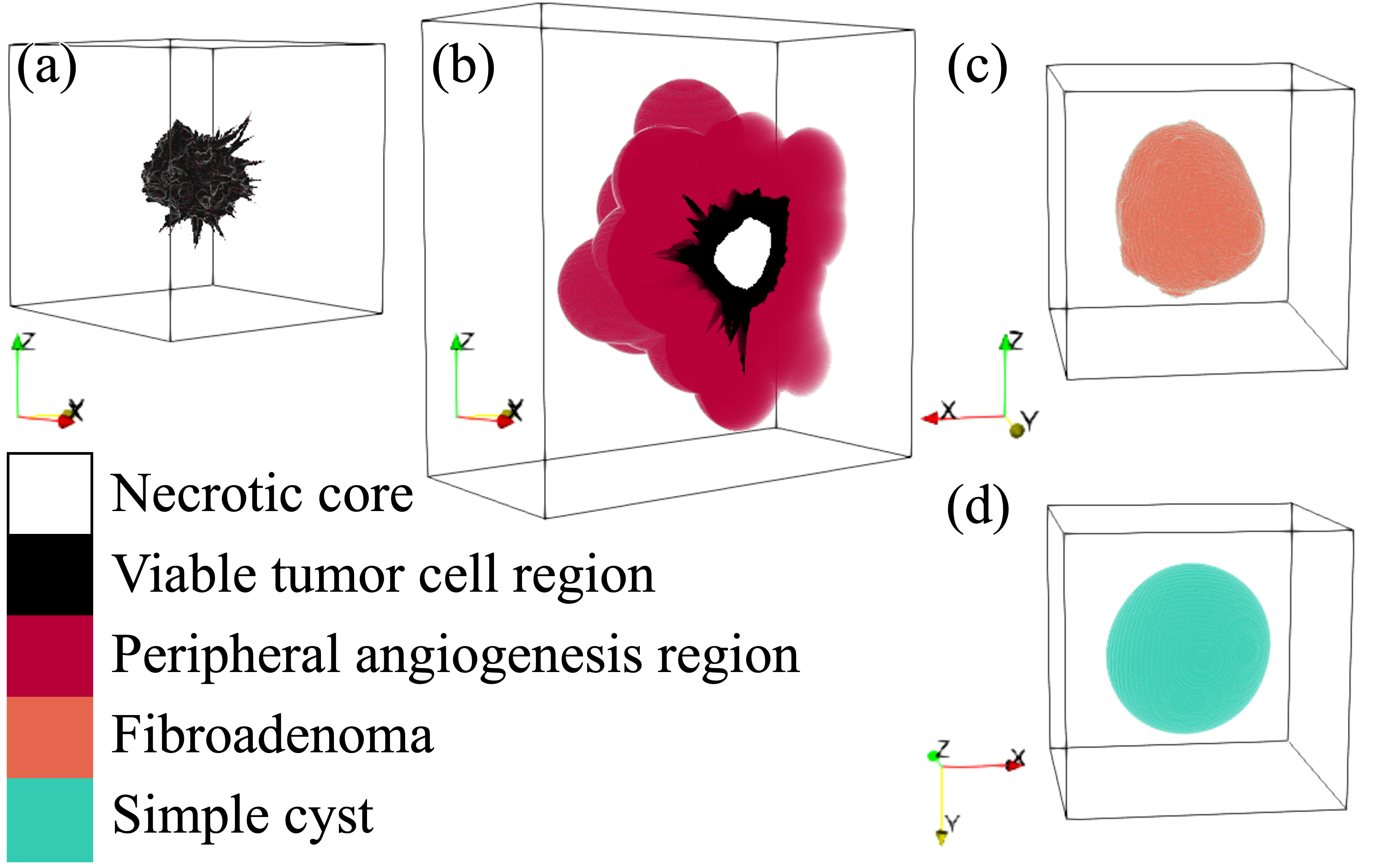}
\caption{Anatomical numerical lesion phantoms (NLPs): (a) malignant tumor with a homogeneous composition (viable tumor cells only), (b) malignant tumor with a heterogeneous composition (viable tumor cells with a necrotic core and a peripheral angiogenesis region), (c) fibroadenoma, and (d) simple cyst. In (b), the lesion volume is clipped along a central plane to reveal the internal composition, whereas the full volumes are displayed in (a), (c), and (d). Volume rendering was performed in ParaView~\cite{ParaView}.}
\label{FIG:2}
\end{figure}

In VICTRE NBP~\cite{victre}, only cancerous masses, for example as shown in Fig.~\ref{FIG:2} (a), and calcifications were modeled. The previous SOA-NBP added two malignant tumor subcomponents, a necrotic core and a peripheral angiogenesis region, resulting in a heterogeneous malignant tumor model, as shown in Fig.~\ref{FIG:2} (b). To enable use of more detailed tumor models, SOA-NBP has been augmented to include benign lesions with functional, optical, and acoustic characteristics distinct from malignant tumors~\cite{Ramala2023, Neuschler2018, Dogan2019, Zhu2005}. Specifically, fibroadenomas, the most common solid masses in women aged 14--35~\cite{Ramala2023}, and simple cysts, the most prevalent breast masses in women aged 35--50~\cite{Berg2010}, were modeled, as illustrated in Figs.~\ref{FIG:2} (c) and (d), respectively.

Both of the considered benign lesions are generally round or oval in shape~\cite{Ramala2023, Santi2025, Hines2010}. Fibroadenomas were modeled with slight surface lobulations to represent their dual composition of glandular and connective tissue, as shown in Fig.~\ref{FIG:2} (c)~\cite{Ramala2023, Santi2025}. In contrast, simple cysts were modeled with smooth contours, consistent with their fluid-filled nature, as presented in Fig.~\ref{FIG:2} (d)~\cite{Hines2010}. Anatomical numerical lesion phantoms (NLPs) for both benign lesions were generated using the VICTRE tool~\cite{victre}, with spicule formation disabled by setting the parameters \texttt{spicule.meanInitial} and \texttt{spicule.stdInitial} to zero in the configuration file. The resulting anatomical NLPs can be inserted at user-specified locations or at candidate sites automatically selected by the VICTRE tool, while avoiding overlap with previously inserted lesions, skin, nipple, or muscle. The definitions of newly introduced or reassigned tissue type labels are described in Appendix~\ref{APPENDIX:B}, and specific lesion shape parameters used in this study are provided in Appendix~\ref{APPENDIX:C}. 

For fibroadenomas, the functional properties and resulting optical absorption coefficient are similar to those of glandular tissue~\cite{Choe2009}. Thus, the functional property values were sampled from the predefined probability distributions for glandular tissue~\cite{Park2023} and assigned to the voxels corresponding to fibroadenomas. The optical absorption coefficient $\mu_a$ was computed from these functional property values following the model in~\cite{Park2023}. Due to the connective tissue component, which exhibits relatively high optical scattering, the optical scattering coefficient $\mu_s$ was calculated using the power law model in \cite{Park2022} with $\mu_s'(\lambda_\text{ref})=1.22$~mm$^{-1}$, $b=1.448$, and $\lambda_\text{ref}=500$~nm~\cite{Jacques2013}. The scattering anisotropy $g=0.96$ and $n=1.36$ were assumed identical to those of glandular tissue~\cite{Park2023}. For acoustic modeling, the sound speed $c$ was drawn from a normal distribution $N(\mu, \sigma)$ with mean $\mu=1.55$~mm/$\mu$s and standard deviation $\sigma=0.032$~mm/$\mu$s~\cite{Li2009}; both the mean and standard deviation values were slightly higher than those of glandular tissue. The remaining acoustic properties were assigned the same values as for glandular tissue. 

Simple cysts consist of nearly pure water, with negligible fat content and no hemoglobin. Accordingly, the water volume fraction $f_w$ was set to 100\%, and the volume fractions of other chromophores, $f_b$, $f_f$, and $f_m$, were set to 0\%. For both optical and acoustic NLPs, the voxels corresponding to this tissue type were assigned the respective properties of water~\cite{Prahl, mua_w, mus_w, engtoolbox, itis, Park2023}.

\subsection{Simulation of the Optoacoustically Induced Initial Pressure Field}
\label{SEC3.2:photoacoustic_effect}
The induced initial pressure distribution $p_{0}(\bm{r}, \lambda)$ was computed according to Eq.~\eqref{eq:1} as the elementwise product of the Gr\"{u}neisen parameter $\Gamma(\bm{r})$, the optical absorption coefficient distribution $\mu_a(\bm{r}, \lambda)$, and the optical fluence distribution $\phi(\bm{r}, \lambda)$. In this study, $\Gamma(\bm{r})$ was set to 1, as often assumed for soft tissue~\cite{fdasimple, Park2023}. The fluence distribution $\phi(\bm{r}, \lambda)$ was simulated with the GPU-accelerated MCX~\cite{mcx1, mcx2}. The optical NBPs were used to specify the optical properties of the medium in the simulation. Most configuration steps were aligned with previous studies employing MCX; however, this framework introduced two novel features. First, a new light source type was introduced to simulate linear segment illuminators used in several 3D OAT imaging systems~\cite{LOUISA-3D, LOIS-3D, TRITOM}. Second, a computationally efficient strategy was adopted to reduce simulation time. Each of these features is described below. 

\begin{figure}
\centering
\includegraphics[width=\columnwidth]{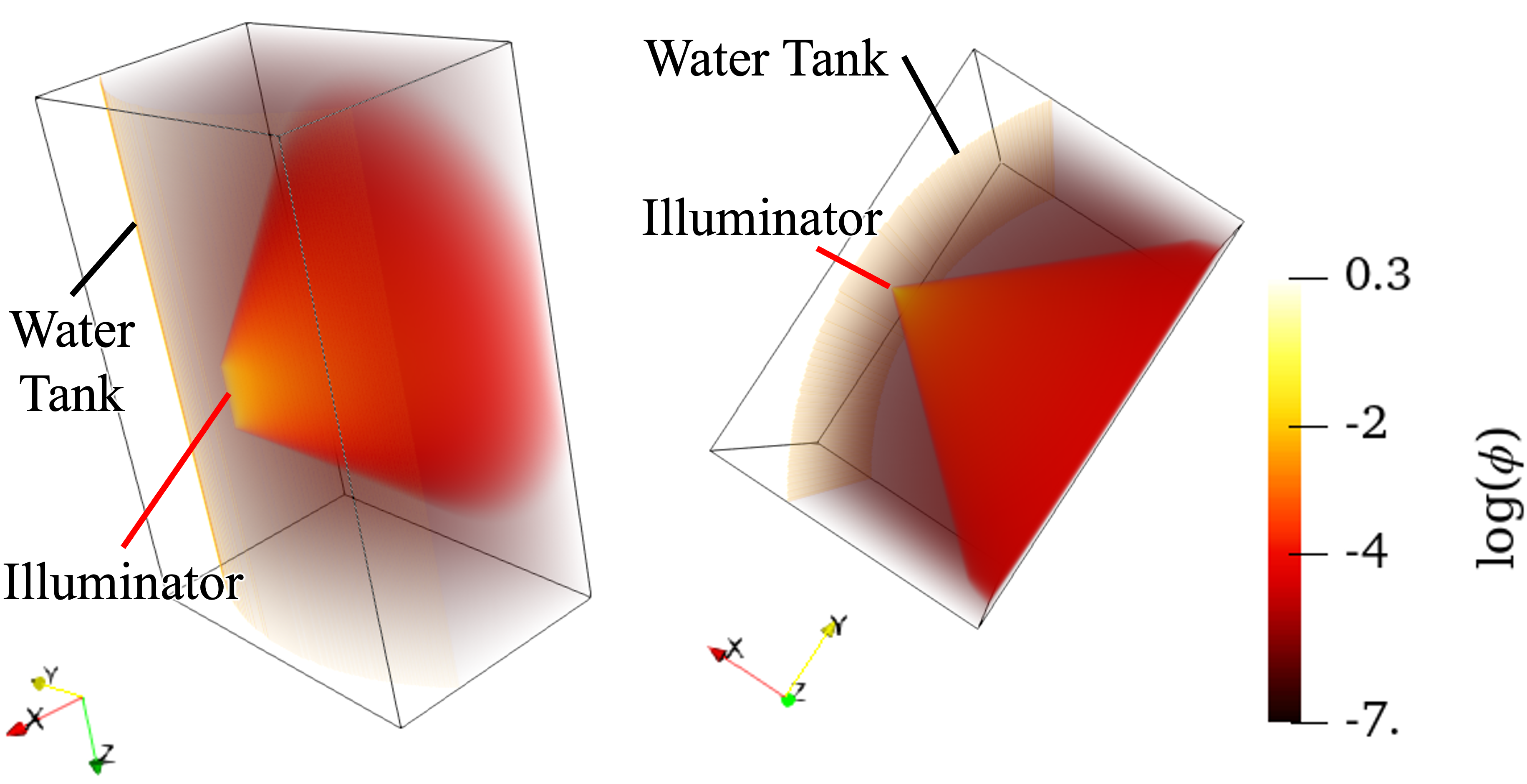}
\caption{Optical fluence distribution $\phi$ simulated using a newly implemented linear segment illuminator, approximated as a continuous line emitter with a conical angular distribution, shown from two different viewing angles. A 20~mm-long illuminator with a numerical aperture of $0.89$ was placed on the surface of the water tank. Volume rendering was performed in ParaView~\cite{ParaView} with logarithmic scaling and value-dependent opacity adjustment applied for effective visualization. \vspace{-1.2em}}
\label{FIG:3}
\end{figure}

Light sources should be configured to closely reflect the light delivery subsystem of the target OAT imager. For systems employing a bottom-up illuminator with divergent emission~\cite{PAM3, Kruger2013}, illumination can be modeled as a cone beam configuration, available as a built-in option in MCX. In this case, the beam position and orientation, together with the cone half-angle determined by the numerical aperture of the illuminator, should be specified in the simulation. As an extension of the proposed framework, systems equipped with linear fiber-optic illuminators~\cite{LOUISA-3D, LOIS-3D, TRITOM} were modeled using a newly implemented source type, a line beam with conical angular emission. In this source type, photon launch positions are sampled uniformly along a predefined optical line illuminator, and their initial propagation directions are sampled from a cone-shaped angular distribution around the specified source axis. The cone half-angle determines the divergence of the emitted photons. The angular distribution is chosen to be uniform in polar angle between 0 and the specified half-angle, with azimuthal angle uniformly distributed in $[0, 2\pi)$. This model approximates the behavior of a densely packed linear emitter array by treating it as a continuous line emitter with controlled divergence. An example optical simulation result from this configuration is shown in Fig.~\ref{FIG:3}.

In the standard approach, $\phi(\bm{r}, \lambda)$ is simulated on the same grid as $p_0(\bm{r}, \lambda)$. For stable computation, the number of photons must greatly exceed the number of voxels in the simulation domain. Consequently, the computation time depends on both the voxel size and domain size. To improve computational efficiency, an approximate strategy was adopted. The optical property distributions $\mu_a(\bm{r}, \lambda)$ and $\mu_s(\bm{r}, \lambda)$ were first downsampled to a coarser grid using trilinear interpolation before the MCX simulation. The resulting $\phi(\bm{r}, \lambda)$ was then upsampled back to the original resolution, also via trilinear interpolation, after which $p_{0}(\bm{r}, \lambda)$ was obtained by multiplying this upsampled $\phi(\bm{r}, \lambda)$ with $\mu_a(\bm{r}, \lambda)$ on the final grid. This process introduces a smoothing effect similar to that applied in the acoustic simulation stage, and therefore the approximation has limited impact on the simulated detected pressure signals.

\begin{figure}
\centering
\includegraphics[width=1\columnwidth]{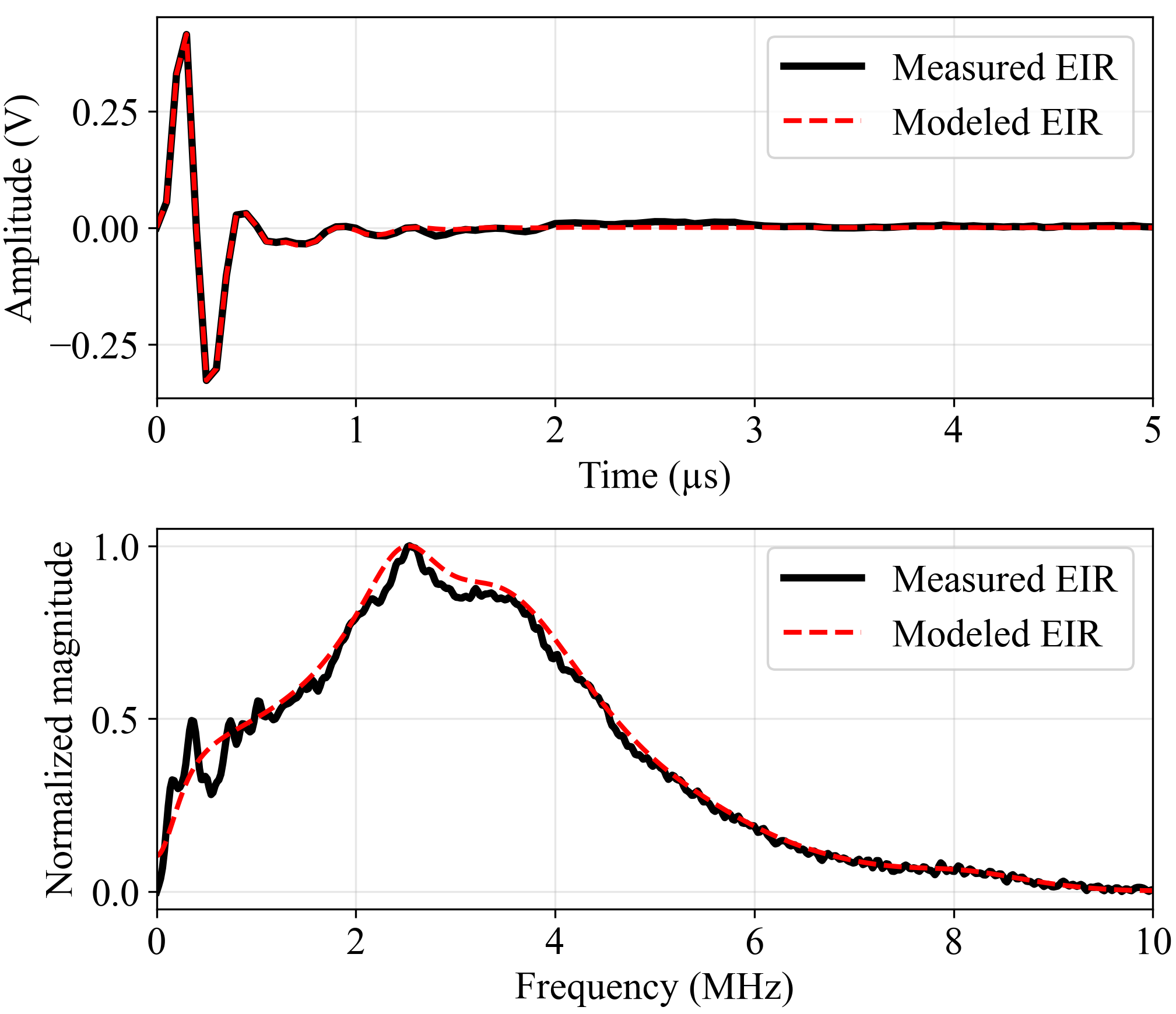}
\caption{Experimentally measured (black solid line) and modeled EIR (red dashed line) of a piezoelectric transducer (Imasonic, France). The time-domain response is shown at the top and the normalized frequency-domain magnitude at the bottom. The measured EIR (black solid line) was acquired by TomoWave Laboratories (Houston, TX), and the modeled EIR (red dashed line) was obtained employing the complex exponential method~\cite{Tallavo2011}, with the model order $\hat{N}=14$ selected based on AIC~\cite{Tallavo2011} (see Appendix~\ref{APPENDIX:D} for details). \vspace{-1.5em}}
\label{FIG:4}
\end{figure}

\subsection{Simulation of Optoacoustic Measurements} 
\label{SEC3.3:optoacoustic_measurements}
The simulation of optoacoustic measurements was carried out using both the initial pressure distribution $p_{0}(\bm{r}, \lambda)$ obtained in Section~\ref{SEC3.2:photoacoustic_effect} and the acoustic NBPs generated with the SOA-NBP tool~\cite{Park2023} as inputs. As outlined in Fig.~\ref{FIG:1}, the first step involves acoustic wave propagation and detection, in which pressure waves from the initial pressure distribution are propagated through the acoustic medium defined by the acoustic NBPs and recorded on finite-sized detection surfaces of the transducers while accounting for their SIR. In the second step, these detected pressure signals are converted into optoacoustic measurements, i.e., voltage signals, by incorporating the EIR of the transducers. 

For the simulation of acoustic wave propagation and detection, a custom Python library was implemented, extending j-Wave~\cite{j-wave}. Two key features that are currently unavailable in j-Wave were incorporated. First, acoustic attenuation described by a frequency power law was implemented. Second, for finite-sized transducers, numerical integration of the simulated pressure field over each transducer's detection surface~\cite{wise2019representing, javaherian2025full} at each time step was incorporated. By accounting for the transducer's SIR, this extension enabled finite-sized transducer modeling and enhanced the realism of the simulated detected pressure signals, albeit within inherent approximations. 

The simulated detected pressure signals were convolved with the EIR of the transducers to obtain synthetic optoacoustic measurements. Any EIR can be incorporated into the framework, whether directly measured or modeled. The EIR can be experimentally measured using a delta-pulse source, where a short laser pulse irradiates a highly absorbing planar surface and generates a pressure pulse~\cite{Conjusteau2009,Conjusteau2010}. However, the measured EIR may include artifacts introduced by the measurement environment, and thus may not fully represent the intrinsic transducer response. To address this, the transducer response was parameterized using the complex exponential method~\cite{Tallavo2011}, which captures the damped resonant behavior of piezoelectric ultrasonic transducers. The model parameters were estimated from the measured EIR using least squares, and the model order was selected using Akaike's information criterion (AIC)~\cite{Tallavo2011}. Further details of the parameterization are provided in Appendix~\ref{APPENDIX:D}. Figure~\ref{FIG:4} shows an example of an experimentally measured EIR and the corresponding modeled EIR. 

\begin{figure*}
	\centering
	\includegraphics[width=\textwidth]{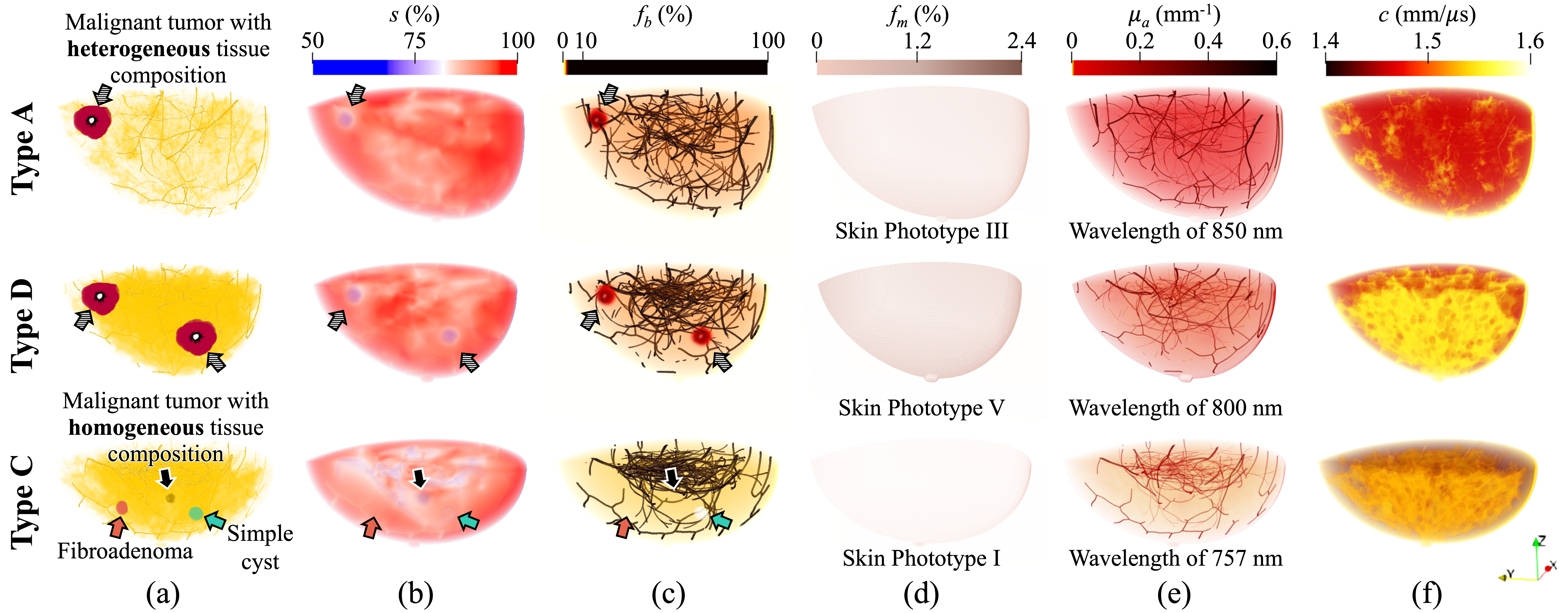}
	\caption{Representative distributions of functional, optical, and acoustic properties: (a) tissue labels, (b) oxygen saturation $s$, (c) blood volume fraction $f_b$, (d) melanosome volume fraction $f_m$, (e) optical absorption coefficient $\mu_a$, and (f) sound speed $c$ of type A, D, and C breasts (top to bottom). The type A breast (top) is naturally shaped, has skin phototype III ($f_m=8.575$\%), and contains one malignant tumor with heterogeneous tissue composition. The type D breast (center) is naturally shaped, has skin phototype V ($f_m=15.25$\%), and contains two malignant tumors with heterogeneous tissue composition. The type C breast (bottom) has a shape constrained by a hemiellipsoidal stabilizer cup and skin phototype I ($f_m=1.9$\%) and includes four lesions: a malignant tumor with homogeneous tissue composition, a malignant tumor with heterogeneous tissue composition, a fibroadenoma, and a simple cyst. All lesions were positioned at $x=0$ except for the two malignant tumor in the type C breast. To visualize internal structures, only half of each breast volume is shown. The malignant tumor with heterogeneous tissue composition in the type C lies outside the displayed field of view. In panel (a), lesion tissues are illustrated with the same colors as in Fig.~\ref{FIG:2}, while healthy tissues are shown in yellow. ParaView~\cite{ParaView} was used for volume rendering. \vspace{-1.2em}}
	\label{FIG:5}
\end{figure*}
\begin{figure}
	\centering
	\includegraphics[width=\columnwidth]{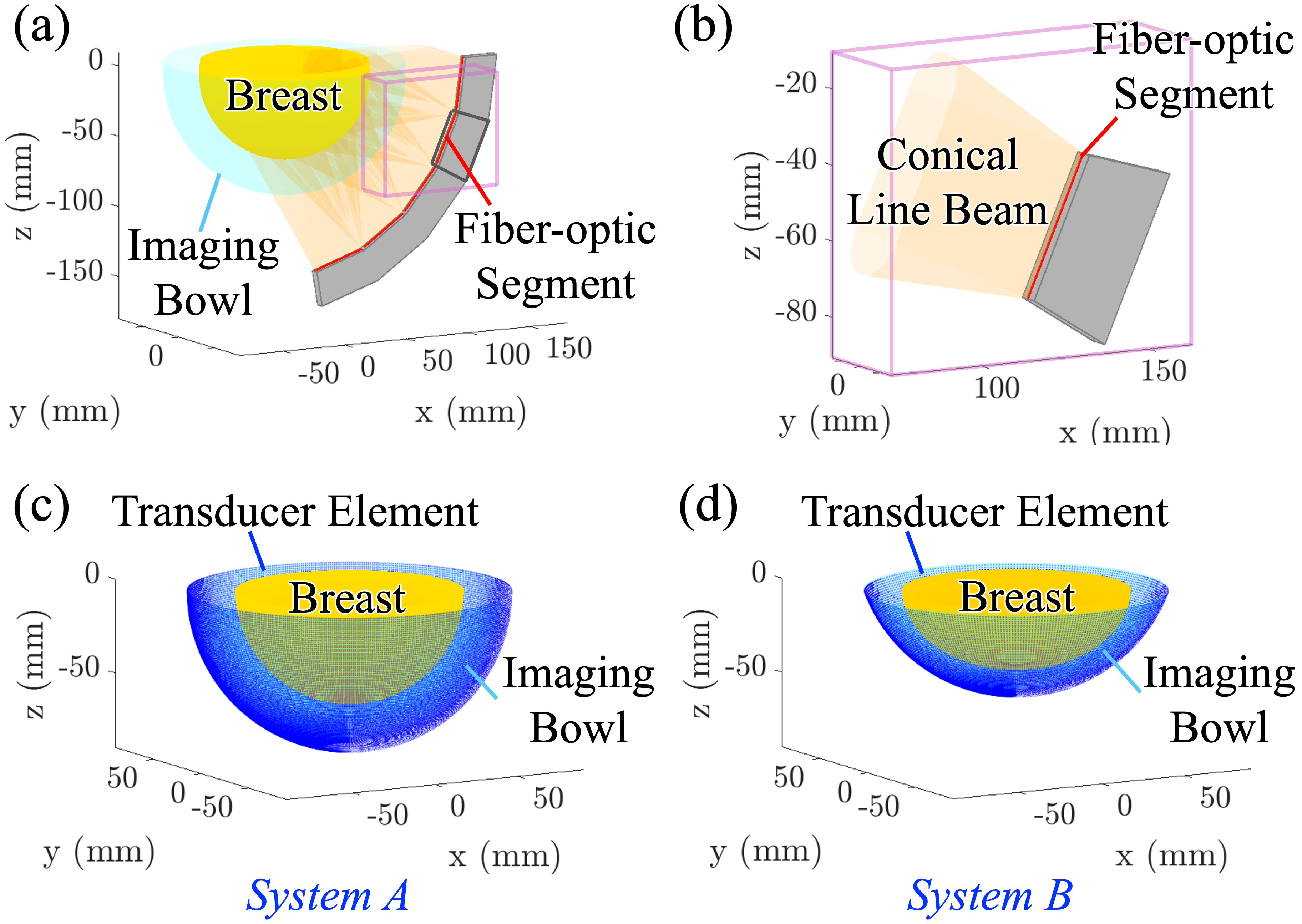}
	\caption{Virtual light delivery subsystem (a, b) and OAT data acquisition subsystem of System A (c) and System B (d). Both systems employed the identical virtual light delivery subsystem, with five linear fiber-optic segments (red lines) attached to the surface of each arc-shaped illuminator, each emitting a line beam with conical angular distribution. A total of 20 illuminator arcs were used, resulting in 100 line beams illuminating the breast. Panels (c) and (d) show the measurement geometries of the virtual OAT data acquisition subsystems in Systems A and B, respectively, including all transducer elements and tomographic views. Each transducer element (blue dot) was oriented toward the center of curvature of the spherical aperture. \vspace{-1.4em}}
	\label{FIG:6}
\end{figure}

\section{Examples of Generated NBPs and Corresponding OAT Images}
\label{SEC4:examples}
To demonstrate the capabilities of the proposed framework, example NBPs are presented in Section~\ref{SEC4.1:example_nbps}. Using the virtual imaging system configuration described in Section~\ref{SEC4.2:system_config} and the simulation settings detailed in Section~\ref{SEC4.3:simulation_config}, OAT measurements were simulated from these example NBPs. The corresponding synthetic OAT images are presented in Section~\ref{SEC4.4:oat_images}. While Section~\ref{SEC3:vi_framework} focused on the extended features introduced in the proposed framework, Section~\ref{SEC4.3:simulation_config} specifies the remaining conventional simulation parameters used in this example, providing more details of the simulation setup to ensure reproducibility.

Although incorporating the transducers' SIR and EIR enables high-fidelity physics modeling in virtual imaging studies, these extended features remain optional depending on research objectives. The examples presented in this section highlight the framework's flexibility by demonstrating datasets generated with both point-like and finite-size transducer assumptions.

\subsection{Example NBPs}
\label{SEC4.1:example_nbps}
Figure~\ref{FIG:5} presents example NBPs, including two naturally shaped breasts of different sizes (first two rows) and one breast constrained in shape by the use of a virtual breast stabilizer cup (bottom row). All NBPs were discretized at a voxel size of 0.125~mm. They incorporate variations across the BI-RADS breast density types~\cite{bi-rads} (A, C, and D), skin phototypes (I, III, and V), and lesion types (a malignant tumor with a homogeneous tissue composition, a malignant tumor with a heterogeneous tissue composition, a fibroadenoma, and a simple cyst). Representative distributions of functional, optical, and acoustic properties are illustrated rather than the complete property sets defined in the NBPs.

\subsection{Virtual Imaging System Configuration}
\label{SEC4.2:system_config}
The NBPs in Fig.~\ref{FIG:5} were virtually imaged using two imaging system configurations. Both systems employed an identical light delivery subsystem but differed in their OAT data acquisition subsystems, with one using point-like transducers and the other finite-size transducers.

The virtual light delivery subsystem was configured with 20 arc-shaped illuminators, each with a radius of 145~mm and a central angle of 80$^\circ$, uniformly distributed in azimuth. As shown in Fig.~\ref{FIG:6} (a), five linear fiber-optic segments were mounted on the surface of each illuminator. A single segment is illustrated in Figure~\ref{FIG:6} (b). Each segment emitted a line beam with a conical angular distribution characterized by a half-angle of 12.5$^\circ$, directed along the axis perpendicular to the segment and oriented toward the origin of the coordinate system, as depicted in Fig.~\ref{FIG:6} (a). 

The virtual OAT data acquisition subsystem of System A was equipped with a hemispherical imaging bowl of radius of 85~mm, filled with water for acoustic coupling. As shown in Fig.~\ref{FIG:6} (c), a rotating arc-shaped transducer array was positioned on the bowl surface, consisting of 108 point-like transducer elements evenly distributed along an arc aperture with a central angle of 80$^\circ$ and a polar angular spacing of approximately 0.75$^\circ$. The rotating arc array completed 480 tomographic views with an azimuthal angular step of 0.75$^\circ$. Each virtual transducer element recorded 3,720 time samples at a sampling frequency of 20~MHz. The modeled EIR shown in Fig.~\ref{FIG:4} was assumed for all elements. 

In System B, the virtual OAT data acquisition subsystem employed a spherical-cap imaging bowl with a radius of 85~mm and a height of 56~mm, as shown in Fig.~\ref{FIG:6} (d). A rotating arc-shaped transducer array was positioned on the bowl surface, consisting of 82 transducer elements of size 1.1~mm $\times$ 1.1~mm, evenly distributed along an arc aperture with a central angle of 70$^\circ$ and a polar angular spacing of approximately 0.86$^\circ$. The rotating arc array stopped 320 tomographic view steps with an azimuthal angular step of 1.125$^\circ$. Each virtual transducer element recorded 2,267 time samples at a sampling frequency of 20~MHz. The same modeled EIR as in System A was assumed.

\subsection{Simulation Configuration}
\label{SEC4.3:simulation_config}
The MCX simulation was performed with the light delivery subsystem configuration described in Section~\ref{SEC4.2:system_config}. The custom line source type implemented in this study was employed to mimic the beams produced by each fiber bundle segment. The optical wavelengths assumed were 757~nm, 800~nm (isosbestic points of deoxy- and oxy-hemoglobin), and 850~nm. Spatially heterogeneous distributions of $\mu_a(\bm{r}, \lambda)$ and $\mu_s(\bm{r}, \lambda)$ were prescribed in MCX using the \texttt{muamus\_float} optical medium format (option 100). The anisotropy factor $g$ and refractive index $n$ were fixed at constant values, using their average values within the breast, since they do not vary significantly across the tissues that constitute most of the breast volume~\cite{Park2023}. Following the strategy introduced in Section~\ref{SEC3.2:photoacoustic_effect} to reduce simulation times, the optical fluence distribution $\phi(\bm{r}, \lambda)$ was simulated on a grid with a voxel size 0.5~mm by employing downsampled optical NBPs, resulting in a simulation domain of 340 $\times$ 340 $\times$ 170 voxels. Each of the 100 line beams was simulated with $10^{8}$ photons for a time duration of 50 ns. The resulting $\phi(\bm{r}, \lambda)$ was then interpolated back to the original 680 $\times$ 680 $\times$ 340 voxel grid (voxel size of 0.25~mm) to compute the initial pressure distribution $p_{0}(\bm{r}, \lambda)$.

The optoacoustic measurement simulation for the type A and D breasts was performed using the data acquisition subsystem of System A, whereas the type C breast was simulated with System B (see Section~\ref{SEC4.2:system_config}). The acoustic NBPs were downsampled from a voxel size of 0.125~mm to 0.25~mm using trilinear interpolation, and the resulting resolution was used in this simulation. For System A, point-like transducer elements were positioned on a grid of size 700 $\times$ 700 $\times$ 350 voxels. After removing duplicate positions introduced by discretization of Cartesian coordinates with the given grid spacing (voxel size of 0.25~mm), 51,472 unique transducer locations remained. This example corresponds to an on-grid approximation of transducer positions for computational simplicity, although the framework is flexible and can also accommodate arbitrary off-grid positions in a manner similar to finite-size transducer modeling~\cite{wise2019representing}. For System B, the SIR of finite-sized transducers was incorporated as described in Section~\ref{SEC3.3:optoacoustic_measurements}, using a simulation grid of size 696 $\times$ 696 $\times$ 288 voxels. To prevent undesired acoustic reflections at the simulation boundaries, a perfectly matched layer was applied. For both systems, the EIR of the transducers was modeled as described in Section~\ref{SEC3.3:optoacoustic_measurements}. 

Electronic noise was modeled as independent and identically distributed Gaussian noise with a mean value of zero. The variance was set based on the measurement data simulated using the type C NBP to yield an SNR of 5 dB. The same noise model was applied to generate the OAT measurements of the other example NBPs. While larger transducers generally provide higher SNR due to spatial averaging of the pressure field, in these examples, however, the same noise level was assumed for both systems as a modeling choice. 

\begin{figure*}
\centering
\includegraphics[width=\textwidth]{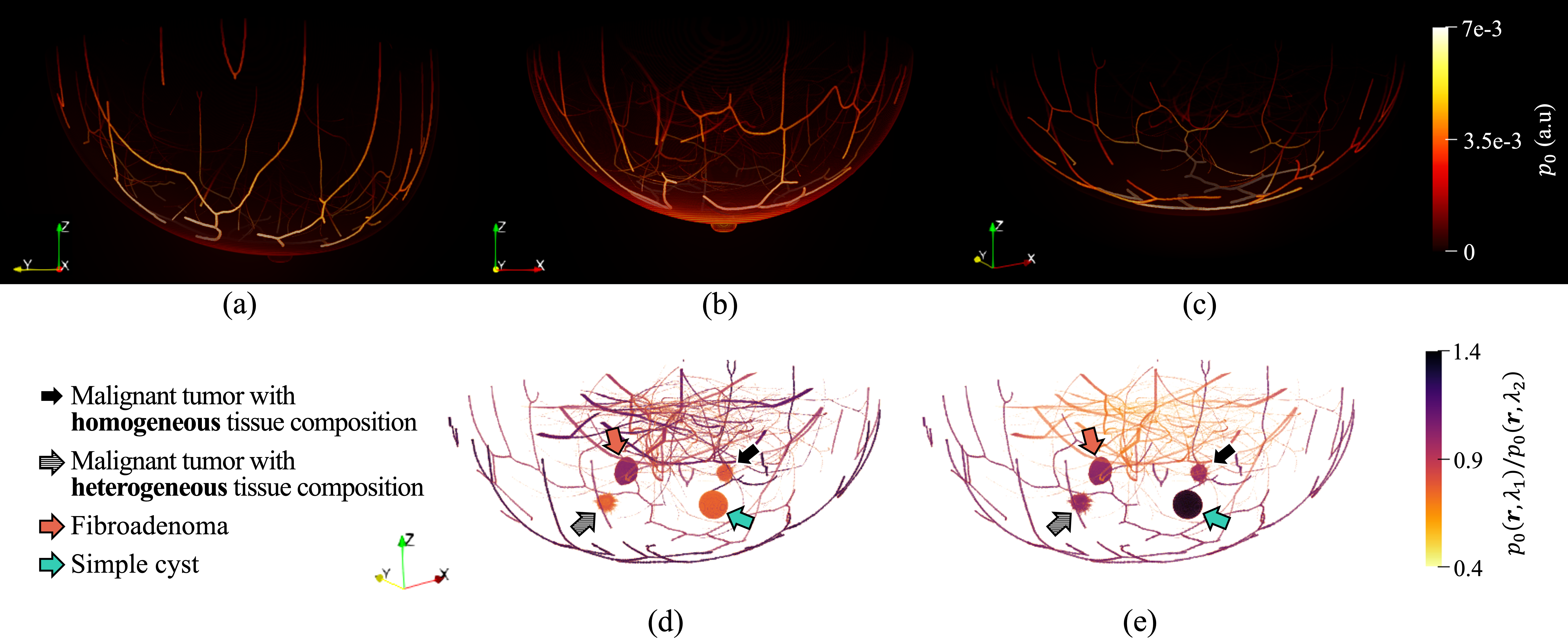}
\caption{Simulated initial pressure distribution $p_0$. Panels (a)-(c) show the $p_0$ distributions of the type A, D, and C breasts from Fig.~\ref{FIG:5} at an illumination wavelengths of $850$, $800$, and $757$~nm, respectively. Panels (d) and (e) show $p_0$ ratios within blood vasculature and lesions of the type C breast, calculated for the wavelength pairs $(\lambda_1, \lambda_2)=(800, 757)$~nm and $(850, 800)$~nm, respectively. Volume rendering was performed in ParaView~\cite{ParaView}. \vspace{-1em}}
\label{FIG:7}
\end{figure*}

\begin{figure}
\centering
\includegraphics[width=\columnwidth]{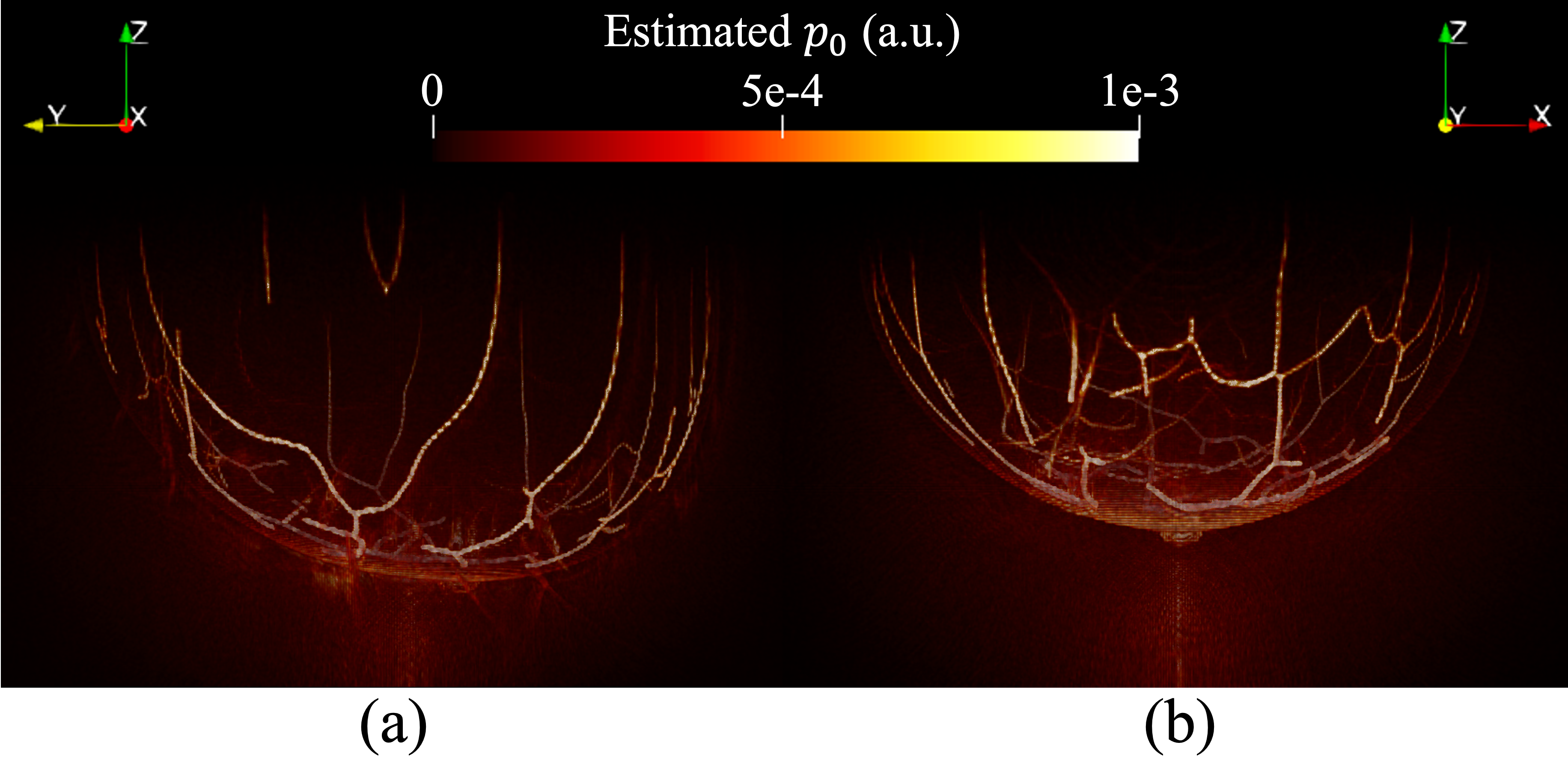}
\caption{Reconstructed OAT images of (a) the type A and (b) the type D breasts from Fig.~\ref{FIG:5}, obtained using UBP with a two-region sound speed model. The corresponding true $p_0$ distributions are shown in Figs.~\ref{FIG:7} (a) and (b), respectively. Volume rendering was performed in ParaView~\cite{ParaView}, with the color map adjusted for improved visibility. In panel (a), the type A breast with skin phototype III shows a skin boundary recognizable through subcutaneous vasculature, whereas in panel (b), the type D breast with skin phototype V, which has higher optical absorption by melanin than phototype III, exhibits a more distinct skin boundary itself. Because point-like transducers were employed in virtual imaging, no blurring effects related to transducer size were observed in both images. \vspace{-1em}}
\label{FIG:8}
\end{figure}

\subsection{Simulated Initial Pressure Distributions and OAT Images}
\label{SEC4.4:oat_images}
Figures~\ref{FIG:7} (a)--(c) show the simulated initial pressure distribution $p_0$ of the type A, D, and C NBPs in Fig.~\ref{FIG:5} at an illumination wavelength of 800, 850, and 757~nm, respectively. Figures~\ref{FIG:7} (d) and (e) present the $p_0$ ratios within blood vasculature and lesions of the type C breast, calculated between the illumination wavelengths of 800 and 757~nm and between 850 and 800~nm, respectively. These simulated $p_0$ distributions were then used as inputs to generate detected pressure, from which the corresponding noisy OAT measurements were acquired using the virtual imaging systems described in Section~\ref{SEC4.2:system_config}.

For image reconstruction, Wiener deconvolution~\cite{Wiener1949, Sompel2016} was first applied to the simulated OAT measurements to compensate for the transducers' EIR, using the experimentally measured response shown in Fig.~\ref{FIG:4}. From the resulting deconvolved data, 3D distributions of $p_0$ were reconstructed with a voxel size of 0.5~mm using the universal backprojection method (UBP)~\cite{Xu2005}, under the assumption of a two-region, piecewise-constant sound speed model. The sound speed values for water ($c_\text{water}$) and breast tissue ($c_\text{breast}$) were selected via a 2D grid search over $[1.48, 1.535]^2$~mm/$\mu$s in increments of 0.005~mm/$\mu$s, choosing the pair that minimized the mean squared error (MSE) between the UBP reconstruction and the true $p_0$ distribution. It should be noted that this strategy is not applicable in practice, since the true $p_0$ distribution is unavailable. Here, this approach was adopted solely for the purpose of presenting example OAT images. Figures~\ref{FIG:8} (a) and (b) show the reconstructed images of the type A breast ($c_\text{water}=1.53$~mm/$\mu$s and $c_\text{breast}=1.535$~mm/$\mu$s) and the type D breast ($c_\text{water}=1.525$~mm/$\mu$s and $c_\text{breast}=1.53$~mm/$\mu$s), respectively. An example OAT image of the type C breast is presented as part of the case study in Section~\ref{SEC5:case_study}. 

\begin{figure*}
	\centering
	\includegraphics[width=\textwidth]{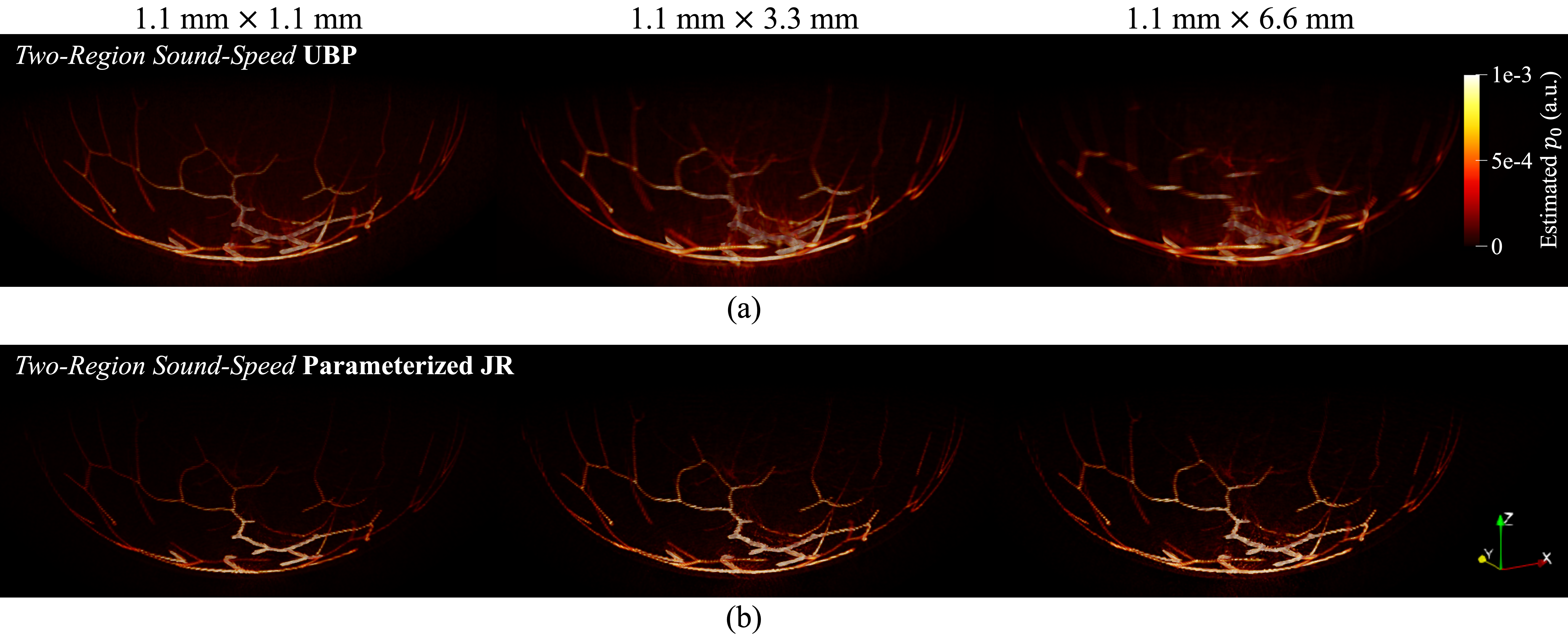}
	\caption{Reconstructed OAT images of the type C breast from Fig.~\ref{FIG:5} using (a) two-region UBP and (b) parameterized JR method incorporating SIR of the transducer. For both methods, the transducer's EIR was identically compensated by preprocessing all simulated measurements with Wiener deconvolution. From left to right, results are shown for System B with transducer element sizes of 1.1~mm $\times$ 1.1~mm, 1.1~mm $\times$ 3.3~mm, and 1.1~mm $\times$ 6.6~mm. The corresponding true initial pressure distribution is provided in Fig.~\ref{FIG:7} (c). Volume rendering was performed in ParaView~\cite{ParaView}, with the color map adjusted but kept identical across all images for consistent visibility. The UBP reconstruction results in (a) exhibited greater resolution degradation as the transducer element size increased, whereas the images estimated using parameterized JR in (b) maintained relatively consistent spatial resolution across all three system designs. These results suggest that, under the given measurement geometry, system designs with transducer elements larger than 1.1~mm $\times$ 3.3~mm require incorporating the transducer's SIR during image reconstruction to achieve high spatial resolution. \vspace{-1.4em}}
	\label{FIG:9}
\end{figure*}

\section{Case Study: Evaluation of OAT Imaging System Designs}
\label{SEC5:case_study}
To demonstrate the utility of the proposed virtual imaging framework, a case study was conducted to evaluate and compare three OAT imaging system designs that employed transducers of varying dimensions. System B that employed transducer elements measuring 1.1~mm $\times$ 1.1~mm, as described above, served as the baseline design. Two additional variants of System B were considered, with enlarged element sizes of 1.1~mm $\times$ 3.3~mm and 1.1~mm $\times$ 6.6~mm.

For each system configuration, virtual imaging was performed using the type C NBP shown in Fig.~\ref{FIG:5} with the simulation setup described in Section~\ref{SEC4.3:simulation_config}. For the systems with 1.1~mm $\times$ 3.3~mm and 1.1~mm $\times$ 6.6~mm elements, the variance of the Gaussian noise was scaled to be inversely proportional to the transducer detection surface area. This scaling reflects area averaging of uncorrelated noise at larger apertures. The resulting SNRs of the simulated pressure data were 8.34 dB and 9.54 dB, respectively.

Two image reconstruction methods were employed: the UBP~\cite{Xu2005} with a two-region sound speed model and a parameterized joint reconstruction (JR) approach~\cite{Matthews2018, Jeong2025}. In the JR, both the initial pressure distribution $p_0$ and the breast sound speed $c_\text{breast}$ were estimated simultaneously, with the sound speed restricted to a two-region parameterization~\cite{Matthews2018}. A single constant value $c_\text{breast}$ was assigned to the breast region, while the water sound speed was fixed at $c_\text{water}=1.525$~mm/$\mu$s. The JR problem was formulated as an optimization-based inverse problem~\cite{Jeong2025}, with the wave equation in Eq.~\eqref{eq:3} modeling the forward propagation of acoustic waves. The model was numerically implemented using the $k$-space pseudo-spectral time-domain method, with the transducers' SIR explicitly incorporated. Non-negativity and total variation constraints were imposed on the $p_0$ distribution. During reconstruction, the acoustic attenuation parameters ($\alpha_0$ and $y$) and density were fixed to known constants for water, whereas for the breast region they were determined based on the generated NBP ensemble. The optimization algorithm, including the constraints and stopping rule for the iterations, followed the approach in~\cite{Jeong2025} and was combined with the two-region parameterization in~\cite{Matthews2018}, both adapted from 2D to the present 3D setting. For comparison, in the two-region UBP, $c_\text{breast}$ was set to the parameterized JR estimate (approximately $1.498$~mm/$\mu$s), and $c_\text{water}$ was fixed to $1.525$~mm/$\mu$s.

The reconstructed images had a voxel size of 0.5~mm, which was larger than the 0.25~mm grid spacing employed in the OAT measurement simulations. This choice was made to avoid the so-called inverse crime~\cite{Kaipio2007}, that is, the use of the same discretization in both the forward model and the reconstruction. The comparison between the two-region UBP and the parameterized JR aimed to investigate the necessity of accounting for the transducer's SIR during reconstruction under the assumed system designs. Although the transducer's EIR could be incorporated into the imaging model at each iteration of the parameterized JR, it was instead compensated identically for both methods in this study by preprocessing all simulated measurements with Wiener deconvolution, as described in Section~\ref{SEC4.4:oat_images}.

Figures~\ref{FIG:9} (a) and (b) present the reconstruction results obtained with the two-region UBP and the parameterized JR, respectively. As the transducer element size increased, the UBP results exhibited pronounced blurring, as shown in Fig.~\ref{FIG:9} (a). In contrast, the parameterized JR maintained relatively consistent spatial resolution across all three system designs, as presented in Fig.~\ref{FIG:9} (b). These results indicate that, under the given measurement geometry, analytic reconstruction methods that do not account for the SIR suffer noticeable resolution degradation when applied to data acquired with transducer elements larger than 1.1~mm $\times$ 3.3~mm. In such cases, incorporating the transducer's SIR during image reconstruction is necessary for achieving high spatial resolution. 

A comprehensive evaluation of imaging system design requires systematic investigations that extend beyond this illustrative case study. Potential directions include: (1) assessing the trade-off between SNR and spatial resolution when enlarging the transducer detection surface, (2) quantifying estimation accuracy and imaging depth when applying JR methods that account for acoustic heterogeneity and attenuation, and (3) evaluating figures of merit for specific clinical tasks, such as signal detection and lesion localization, rather than relying solely on conventional image quality measures. These investigations can be effectively pursued by use of the proposed virtual imaging framework.

\section{Conclusion}
\label{SEC6:conclusion}
In this work, a comprehensive framework was established to enable virtual imaging studies of 3D qOAT of the breast. Building upon the previously proposed SOA-NBP tool, the framework incorporates a two-layer skin model that captures variations in skin tone. It also extends the lesion modeling by including two benign lesion types, fibroadenomas and simple cysts, in addition to malignant tumors. By use of stochastically generated NBPs, the framework enables end-to-end simulation of the entire imaging process, from optoacoustically induced initial pressure generation to acoustic wave propagation and optoacoustic measurement acquisition. Transducer characteristics, including both the SIR and EIR, were explicitly modeled to faithfully replicate imaging physics under realistic system conditions.

To accelerate further research, the datasets generated in this study are publicly available at \emph{[link to be provided upon acceptance of the manuscript]}. The framework is broadly applicable, with potential use cases extending beyond OAT. For instance, NBPs generated employing the SOA-NBP tool have already been utilized in optical imaging studies~\cite{Hossain_2025}. The present framework further provides a platform for systematic investigations of system designs and reconstruction methods in both optical and acoustic inversion. Future studies could further extend the framework by incorporating lesion-dependent vascular features, such as the distinct spatial distribution and connectivity of blood vessels that are typically observed in OAT images around malignant lesions compared with benign lesions.

\section*{Declaration of Competing Interest}
The authors declare no known competing financial interests or personal relationships that could have appeared to influence the work reported in this paper.

\section*{Acknowledgments}
This work was supported in part by the National Institutes of Health, United States grants EB031585 and EB034261. This work used the Delta system at the National Center for Supercomputing Applications through allocation MDE230007 from the Advanced Cyberinfrastructure Coordination Ecosystem: Services \& Support (ACCESS) program, which is supported by U.S. National Science Foundation grants \#2138259, \#2138286, \#2138307, \#2137603, and \#2138296. The authors would like to thank Drs. Mohammad Eghtedari and Andre Kajdacsy-Balla for their consultation on the numerical modeling of lesions to ensure realistic representation of their pathological and functional properties. They are also grateful to Dr. Alexander A. Oraevsky for insightful discussions on EIR modeling and for providing EIR measurement data, and to Dr. Richard Su for collecting the EIR measurement data.

\appendix
\section{Existing Simulation Tools for Virtual Optoacoustic Imaging}
\label{APPENDIX:A}
For high-fidelity simulation of photon transport in biological tissue, several MC simulation tools are available, including GPU-accelerated simulators such as MCX and MMC (implemented in C++ and CUDA) and MATLAB toolboxes such as MCmatlab and Valo MC. MCX and MMC also offer MATLAB wrappers, known as MCXLAB~\cite{mcx1, mcx2} and MMCLAB~\cite{mmc}, respectively. MCX~\cite{mcx1, mcx2} and MCmatlab~\cite{mcmatlab} use Cartesian grids, enabling simulation of light transport in media with voxel-level heterogeneity. MMC~\cite{mmc} and Valo MC~\cite{valo_mc} employ finite tetrahedral elements, supporting modeling of element-level heterogeneity. These tools provide various light source models. ValoMC includes direct, cosine, Gaussian, isotropic, and pencil light source types. MCmatlab supports pencil, isotropic (line or point), infinite plane wave, Laguerre-Gaussian LG01, radial-factorizable (e.g., Gaussian), and X/Y factorizable (e.g., rectangular LED emitter) source models. MCX and MMC provide pencil, isotropic, uniform cone, collimated Gaussian, one-sheeted hyperboloid Gaussian, uniform planar, uniform disk, and patterned source types.

The most widely used acoustic wave simulator in the field of optoacoustic imaging is the MATLAB toolbox k-Wave~\cite{k-wave}, which solves the acoustic wave equation using the $k$-space pseudo-spectral time-domain method. It supports arbitrarily specified (i.e., off-grid) transducer positions to reduce mismatch with actual placement~\cite{Wise2019} and allows modeling of finite-sized transducers by incorporating their SIR. A corresponding Python library, j-Wave~\cite{j-wave}, has also been developed as a JAX-based differentiable acoustic simulator, but it currently lacks acoustic attenuation modeling and support for both off-grid transducer positioning and finite-sized transducer modeling.

\section{Tissue Label Definitions for SOA-NBP}
\label{APPENDIX:B}
Tissue types were labeled with unsigned 8-bit integers. In both the prior SOA-NBP and the present extension, tissue types corresponding to those in the VICTRE NBP~\cite{victre} retained their original label values, whereas newly introduced tissue types were assigned values unused in VICTRE to ensure compatibility with the existing label definitions.

Voxels labeled as skin with value 2 in VICTRE and the previous SOA-NBP were assigned to the dermis-only layer. The epidermis-included layer was re-labeled to value 3, which is unsued in VICRE. In the VICTRE, the tissue label value 200 denotes the cancerous mass; this value was retained for the VTC region in the malignant tumor models of SOA-NBP. The necrotic core, peripheral angiogenesis region, fibroadenoma, and simple cyst were assigned label values of 210, 190, 180, and 160, respectively. These values are unused in VICTRE~\cite{victre}. Tissue type labels are summarized in Table~\ref{tblapp1}, with newly introduced types in SOA-NBP shown in bold. 

\begin{table}[width=.75\linewidth,cols=4,pos=h]
\caption{Tissue type labels}\label{tblapp1}
\begin{tabular*}{\tblwidth}{@{}LL@{}}
\toprule
Tissue type & Label \\ 
\midrule
Fat                 & 1 \\
Dermis-only         & 2 \\
\textbf{Epidermis-included}  & \textbf{3} \\
Glandular           & 29 \\
Nipple              & 33 \\
Ligament            & 88 \\
Terminal duct lobular unit & 95 \\
Duct                & 125 \\
Artery              & 150 \\
\textbf{Simple cyst}         & \textbf{160} \\
\textbf{Fibroadenoma}        & \textbf{180} \\
\textbf{Peripheral angiogenesis} & \textbf{190} \\
Viable tumor cell   & 200 \\
\textbf{Necrotic core}       & \textbf{210} \\
Vein              & 225 \\
\bottomrule
\end{tabular*}
\end{table}

\section{Example Parameters for Lesion Shape Modeling}
\label{APPENDIX:C}
Anatomical NLPs for modeling malignant tumors, fibroadenomas, and simple cysts can be generated using the VICTRE tool with a configuration file~\cite{victre}. In this file, the voxel size parameter \texttt{imgRes} (mm) and the lesion radius parameter \texttt{alpha} (mm) are specified by the users. Table~\ref{tblapp2} lists example VICTRE tool parameter values used in this study to model the VTC region of malignant tumors, fibroadenomas, and simple cysts. These values control the overall geometry, boundary irregularity, and surface characteristics of each lesion type. For the VTC region, most values follow the default VICTRE NLP configuration, except for \texttt{spicule.meanInitial}, \texttt{spicule.stdInitial}, \texttt{spicule.meanInitRad}, \texttt{spicule.stdInitRad}, \texttt{spicule.meanRad- Dec}, \texttt{spicule.stdRadDec}, \texttt{spicule.meanInitLen}, \texttt{spicule.std- InitLen}, and \texttt{spicule.meanLenDec}, which were modified. All other parameters remain identical to the default configuration. 

\begin{table}
\caption{Lesion shape parameters}\label{tblapp2}
\begin{tabular*}{\tblwidth}{@{}LLLL@{}}
\toprule
Parameter & VTC region & Fibroadenoma & Cyst \\ 
\midrule
\texttt{base.complexity}       & 1     & 0.8   & 0.5 \\
\texttt{mass.lMax}             & 4     & 7     & 5 \\
\texttt{mass.meanSigma2}       & 0.31  & 0.2   & 0.03 \\
\texttt{mass.stdSigma2}        & 0.04  & 0.04  & 0.01 \\
\texttt{mass.powerLaw}         & 4     & 5     & 5 \\
\texttt{mass.meanLF}           & 611.2 & 30    & 10 \\
\texttt{mass.stdLF}            & 70.6  & 8     & 3 \\
\texttt{mass.meanShape}        & 0.36  & 0.75  & 1 \\
\texttt{mass.stdShape}         & 0.48  & 0.8   & 0 \\
\texttt{mass.meanLFRad}        & 0.299 & 0.2   & 0 \\
\texttt{mass.stdLFRad}         & 0.073 & 0.06  & 0 \\
\texttt{mass.meanLFLen}        & 0.113 & 0.1   & 0 \\
\texttt{mass.stdLFLen}         & 0.021 & 0.025 & 0 \\
\texttt{mass.meanFuzzAlpha}    & 0.015 & 0.01  & 0 \\
\texttt{mass.stdFuzzAlpha}     & 0     & 0     & 0 \\
\texttt{spicule.meanInitial}   & 100   & 0     & 0 \\
\texttt{spicule.stdInitial}    & 365   & 0     & 0 \\
\texttt{spicule.meanNeigh}     & 8.98  & NA    & NA \\
\texttt{spicule.stdNeigh}      & 1.89  & NA    & NA \\
\texttt{spicule.meanInitRad}   & 0.1   & NA    & NA \\
\texttt{spicule.stdInitRad}    & 0.2   & NA    & NA \\
\texttt{spicule.meanRadDec}    & 0.33  & NA    & NA \\
\texttt{spicule.stdRadDec}     & 0.25  & NA    & NA \\
\texttt{spicule.meanInitLen}   & 0.2   & NA    & NA \\
\texttt{spicule.stdInitLen}    & 0.2   & NA    & NA \\
\texttt{spicule.meanLenDec}    & 0.4   & NA    & NA \\
\texttt{spicule.stdLenDec}     & 0.3   & NA    & NA \\
\texttt{spicule.meanContProb}  & 0.717 & NA    & NA \\
\texttt{spicule.stdContProb}   & 0.057 & NA    & NA \\
\texttt{spicule.meanBranchAng} & 6.55 & NA    & NA \\
\texttt{spicule.stdBranchAng}  & 0.62  & NA    & NA \\
\bottomrule
\end{tabular*}
\end{table}

\section{Complex Exponential Parameterization of the EIR}
\label{APPENDIX:D}
The EIR was modeled employing the complex exponential method~\cite{Tallavo2011}, which represents transient and deterministic signals as a finite sum of complex exponential functions. The model is parameterized by $\boldsymbol{\Theta} = \{a_i, \xi_i, f_i, \varphi_i\}_{i=1}^{N}$, where $N$ is the number of damped cosine components, and $a_i$, $\xi_i$, $f_i$, and $\varphi_i$ denote the amplitude, damping ratio, frequency, and phase of the $i$-th component, respectively. Using this parameterization, the modeled EIR, denoted as $\mathbf{H}^\text{e} \in \mathbb{R}^{L}$ and represented as a discretized time-domain vector of length $L$ with sampling interval $\Delta t$, is expressed as~\cite{Tallavo2011}:
\begin{align}
\label{eq:eir}
&[\mathbf{\mathbf{H}^\text{e}}]_{l}
=\!{h^{e}}(t; \boldsymbol{\Theta})\big|_{t=l \Delta t} \notag \\
&=\!\sum_{i=1}^{N}\!a_i e^{-\xi_i 2 \pi f_i t}\!\cos(2\pi\!f_i t\!+\!\varphi_i)\big|_{t=l \Delta t},\ {\scriptstyle l = 0, \dots, L - 1}.
\end{align}
The parameter set $\boldsymbol{\Theta}$ was estimated from the measured EIR using least squares~\cite{Tallavo2011}. The model order $N$ was selected using AIC to balance model complexity with accuracy. In this study, $\hat{N}$ was chosen as the value of $N$ within the range $[2, 20]$ that minimized AIC~\cite{Tallavo2011}.

\section*{Data Availability}
\label{data}
A total of 1,020 datasets are publicly available via the Illinois Data Bank under the Creative Commons Zero license. Each dataset includes NBPs with embedded NLPs of heterogeneous tissue composition, providing distributions of functional, optical, and acoustic properties, as well as simulated distributions of multi-wavelength optical fluence, initial pressure, and OAT measurements. 

\printcredits

\bibliographystyle{cas-model2-names}

\bibliography{references}



\end{document}